\documentclass[12pt,twoside,a4paper]{article}
\usepackage{amsmath,amssymb,latexsym,theorem,bbm,natbib,epsfig,color,subfigure}
\usepackage{multirow,graphicx,array,rotating}  
\newfont{\msa}{msam10 scaled\magstep1}
\newfont{\ssmsa}{msam9}

\def\crps{\mathop{\hbox{\rm CRPS}}}
\def\crpss{\mathop{\hbox{\rm CRPSS}}}
\def\bs{\mathop{\hbox{\rm BS}}}
\def\bss{\mathop{\hbox{\rm BSS}}}

\numberwithin{equation}{section}

\title{Censored and shifted gamma distribution based EMOS model for probabilistic quantitative precipitation forecasting} 

\author{S. Baran$^{\mathrm{a}}$ and 
    D. Nemoda$^{\mathrm{a,b}}$ \\ [2mm]
{\small
$^{\mathrm a}$Faculty of Informatics, University of Debrecen, Hungary} \\
{\small $^{\mathrm b}$Faculty of Mechanical Engineering and Informatics, 
         University of Miskolc, Hungary}
}

\date{}
\begin{document}
\pagestyle{myheadings}

\maketitle

\begin{abstract}
Recently all major weather prediction centres provide forecast ensembles of different weather quantities which are obtained from multiple runs of numerical weather prediction models with various initial conditions and model parametrizations. However, ensemble forecasts often show an underdispersive character and may also be biased, so that some post-processing is needed to account for these deficiencies. Probably the most popular modern post-processing techniques are the ensemble model output statistics (EMOS) and the Bayesian model averaging (BMA) which provide estimates of the density of the predictable weather quantity.

In the present work an EMOS method for calibrating ensemble forecasts of precipitation accumulation is proposed, where the predictive distribution follows a censored and shifted gamma (CSG) law with parameters depending on the ensemble members. The CSG EMOS model is tested on ensemble forecasts of 24 h precipitation accumulation of the eight-member University of Washington mesoscale ensemble and on the 11 member ensemble produced by the operational Limited Area Model Ensemble Prediction System of the Hungarian Meteorological Service. The predictive performance of the new EMOS approach is compared with the fit of the raw ensemble, the generalized extreme value (GEV) distribution based EMOS model and the gamma BMA method. According to the results, the proposed CSG EMOS model slightly outperforms the GEV EMOS approach in terms of calibration of probabilistic and accuracy of point forecasts and shows significantly better predictive skill that the raw ensemble and the BMA model.

\end{abstract}

\markboth{S. Baran and D. Nemoda}{Gamma EMOS model for probabilistic precipitation forecasting} 
\section{Introduction}
  \label{sec:sec1}

Reliable and accurate prediction of precipitation is of great importance in agriculture, tourism, aviation and in some other fields of economy as well. In order to represent the uncertainties of forecasts based on observational data and numerical weather prediction (NWP) models one can run these models with different initial conditions or change model physics, resulting in a forecast ensemble \citep{leith}. In the last two decades this approach has became a routinely used technique all over the world and recently all major weather prediction centres have their own operational ensemble prediction systems (EPS), e.g. the Consortium for Small-scale Modelling (COSMO-DE) EPS of the German Meteorological Service \citep[DWD; ][]{gtpb,btg}, the
Pr\'evision d'Ensemble ARPEGE (PEARP) EPS of M\'eteo France \citep{dljbac} or the EPS of  the independent intergovernmental  European Centre for Medium-Range Weather Forecasts \citep{ecmwf}. With the help of a forecast ensemble one can estimate the distribution of the predictable weather quantity which opens up the door for probabilistic forecasting \citep{gr05}. By post-processing the raw ensemble the most sophisticated probabilistic methods result in full predictive cumulative distribution functions (CDF) and correct the possible bias and  underdispersion of the original forecasts. The underdispersive character of the ensemble has been observed with several ensemble prediction systems \citep{bhtp} and this property also leads to the lack of calibration. Using predictive CDFs one can easily get consistent estimators of probabilities of various meteorological events or calculate different prediction intervals.

Recently, probably the most widely used ensemble post-processing methods leading to full predictive distributions \citep[for an overview see e.g.][]{gneiting14,wfk} are the Bayesian model averaging \citep[BMA;][]{rgbp} and the non-homogeneous regression or ensemble model output statistics \citep[EMOS;][]{grwg}, as they are partially implemented in the {\tt ensembleBMA} and {\tt ensembleMOS} packages of {\tt R} \citep{frgsb}. 

The BMA predictive probability density function (PDF) of the future weather quantity is the mixture of individual PDFs corresponding to the ensemble members with mixture weights determined by the relative performance of the ensemble members during a given training period. To model temperature or sea level pressure a normal mixture seems to be appropriate \citep{rgbp}, wind speed requires non-negative and skewed component PDFs such as gamma \citep{sgr10} or truncated normal \citep{bar} distributions, whereas for surface wind direction a von Mises distribution \citep{bgrgg} is suggested. 
However, in some situations BMA post-processing might result, for instance, in model overfitting \citep{hamill07} or over-weighting climatology \citep{hsmhs}.

In contrast to BMA, the EMOS technique uses a single parametric PDF with parameters depending on the ensemble members. Again, for temperature and sea level pressure the EMOS predictive PDF is normal \citep{grwg}, whereas for wind speed truncated normal \citep{tg}, generalized extreme value \citep[GEV;][]{lt}, censored logistic \citep{mmzw}, truncated logistic, gamma \citep{schm} and log-normal \citep{bl} distributions are suggested. 

However, statistical calibration of ensemble forecasts of precipitation is far more difficult than the post-processing of the above quantities. As pointed out by \citet{schham}, precipitation has a discrete-continuous nature with a positive probability of being zero and larger expected precipitation amount results in larger forecast uncertainty. \citet{srgf} introduced a BMA model 
where each individual predictive PDF consists of a discrete component at zero and a gamma distribution modelling the case of positive precipitation amounts. \citet{wilks09} uses extends logistic regression to provide full probability distribution forecasts, whereas \citet{sch} suggests an EMOS model based on a censored GEV distribution. Finally, \citet{schham} propose a more complex three step approach where they first fit a censored and shifted gamma (CSG) distribution  model to the climatological distribution of observations, then after adjusting the forecasts to match this climatology derive three ensemble statistics, and with the help of a nonhomogeneus regression model connect these statistics to the CSG model.

Based on the idea of \citet{schham} we introduce a new EMOS approach which directly models the distribution of precipitation accumulation with a censored and shifted Gamma predictive PDF. The novel EMOS approach is applied to 24 hour precipitation accumulation forecasts of the eight-member University of Washington mesoscale ensemble \citep[UWME;][]{em05} and the 11 member operational EPS of the Hungarian Meteorological Service (HMS) called Aire Limit\'ee Adaptation dynamique D\'evelopment International - Hungary EPS \citep[ALADIN-HUNEPS;][]{hkkr,horanyi}. In these case studies the performance of the proposed EMOS model is compared to the forecast skills of the GEV EMOS method of \citet{sch} and to the gamma BMA approach of \citet{srgf} serving as benchmark models.

\section{Ensemble Model Output Statistics}
  \label{sec:sec2}

As mentioned in the Introduction, the EMOS predictive PDF of a future weather quantity is a single parametric distribution with parameters depending on the ensemble members. Due to the special discrete-continuous nature of precipitation one should think only of non-negative predictive distributions assigning positive mass to the event of zero precipitation. Mixing a point mass at zero and a separate non-negative distribution does the job \citep[see e.g. the BMA model of][]{srgf}, but left censoring of an appropriate continuous distribution at zero can also be a reasonable choice. The advantage of the latter approach is that the probability of zero precipitation can directly be derived from the corresponding original (uncensored) cumulative distribution function (CDF), so the cases of zero and positive precipitation can be treated together. The EMOS model of \citet{sch} utilizes a censored GEV distribution with shape parameter ensuring a positive skew and finite mean, whereas our EMOS approach is based on a CSG distribution appearing in the more complex model of \citet{schham}.

\subsection{Censored and shifted gamma EMOS model}
   \label{subs:subs2.1}

Consider a gamma distribution \ $\Gamma (k,\theta)$ \ with shape \ $k>0$ \ and scale \ $\theta>0$ \ having PDF
\begin{equation*}
g_{k,\theta}(x):= \begin{cases}
\frac {x^{k-1}{\mathrm e}^{-x/\theta}}{\theta ^k \Gamma(k)}, &x>0,\\
0, & \text{otherwise,}
\end{cases}
\end{equation*}
where \ $\Gamma(k)$ \ denotes value of the gamma function at \ $k$. \ A gamma distribution can also be parametrized by its mean \ $\mu>0$ \ and standard deviation \ $\sigma>0$ \ using expressions
\begin{equation*}
k=\frac {\mu^2}{\sigma^2} \qquad \text{and} \qquad \theta =\frac {\sigma ^2}{\mu}.
\end{equation*}
Now, let \ $\delta>0$ \ and denote by \ $G_{k,\theta}$ \ the CDF of the \ $\Gamma (k,\theta)$ \ distribution. Then the shifted gamma distribution left censored at zero (CSG) \ $\Gamma^{\,0}(k,\theta,\delta)$ \ with shape \ $k$, \ scale \ $\theta$ \ and shift \ $\delta$ \ can be defined with CDF
\begin{equation}
  \label{eq:CSGCDF}
 G^{\,0}_{k,\theta,\delta}(x):=\begin{cases}
   G_{k,\theta}(x+\delta), & x\geq 0,\\
   0, & x<0.
  \end{cases}
\end{equation} 
This distribution assigns mass \ $G_{k,\theta}(\delta)$ \ to the origin and has generalized PDF
\begin{equation*}
g^{\,0}_{k,\theta,\delta}(x):={\mathbb I}_{\{x=0\}}G_{k,\theta}(\delta)+{\mathbb I}_{\{x>0\}}\big(1-G_{k,\theta}(\delta)\big)g_{k,\theta}(x+\delta),
\end{equation*} 
where \ $\mathbb I_A$ \ denotes the indicator function of the set \ $A$. \ Short calculation shows that the mean \ $\kappa$ \ of  \ $\Gamma^{\,0}(k,\theta,\delta)$ \ equals
\begin{equation*}
  \kappa=\theta k \big(1-G_{k,\theta}(\delta)\big)\big(1-G_{k+1,\theta}(\delta)\big) - \delta \big(1-G_{k,\theta}(\delta)\big)^2,
\end{equation*}
whereas the $p$-quantile \ $q_p$ \ ($0<p<1$) \ of \eqref{eq:CSGCDF} equals \ $0$  \ if \ $p\leq G_{k,\theta}(\delta)$, \ and the solution of \ $G_{k,\theta}(q_p+\delta)=p$, \ otherwise. 

Now, denote by \ $f_1,f_2,\ldots ,f_m$ \ the ensemble of
distinguishable forecasts of precipitation accumulation for a given location
and time. This means that each ensemble member can be identified and
tracked, which holds for example for the UWME (see Section
\ref{subs:subs3.1}) or for the COSMO-DE EPS of the DWD. In the proposed CSG EMOS model the ensemble members are linked to the mean \ $\mu$ \ and variance \ $\sigma^2$ \ of the underlying gamma distribution via equations
\begin{equation}
   \label{eq:Link}
\mu = a_0+a_1f_1+ \cdots +a_mf_m \qquad \text{and} \qquad  \sigma^2 = b_0+b_1\overline f, %\quad \text{with} \quad S^2:=\frac 1{m\!-\!1}\sum_{k=1}^m\big (f_k\!-\!\overline f\big)^2,
\end{equation}
where \ $\overline f$ \ denotes the ensemble mean.
Mean parameters \ $a_0,a_1, \ldots ,a_m\geq 0$ \ and variance parameters
\ $b_0, b_1 \geq 0$ \  of model \eqref{eq:Link} can be estimated from the training data, consisting of ensemble members and verifying observations from the preceding \ $n$ \ days, by optimizing an appropriate verification score 
(see Section \ref{subs:subs2.2}).

However, most of the currently used EPSs produce ensembles containing groups of statistically indistinguishable ensemble members which are obtained with the help of random perturbations of the initial conditions. This is the case for the ALADIN-HUNEPS ensemble described in Section \ref{subs:subs3.2} or for the 51 member ECMWF ensemble. The existence of several exchangeable groups is also a natural property of some multi-model EPSs such as the the THORPEX Interactive Grand Global Ensemble \citep{tigge15} or the GLAMEPS ensemble \citep{iversen11}.

Suppose we have \ $M$ \ ensemble members divided into \ $m$ \ exchangeable groups, where the \ $k$th \ group contains \ $M_k\geq 1$ \ ensemble members, such that \ $\sum_{k=1}^mM_k=M$. \ Further, we denote by \ $f_{k,\ell}$ \ the  $\ell$th member of the $k$th group. In this situation ensemble members within a given group should share the same parameters \citep{gneiting14} resulting in the exchangeable version 
\begin{equation}
   \label{eq:Linkex}
\mu = a_0+a_1\sum_{\ell_1=1}^{M_1}f_{1,\ell_1}+ \cdots
  +a_m\sum_{\ell_m=1}^{M_m} f_{m,\ell_m}, \qquad 
 \sigma^2 =b_0+b_1\overline f,
\end{equation}
of model \eqref{eq:Link}. 

Note, that the expression of the mean (or location) as an affine function of the ensemble is general in EMOS post-processing \citep[see e.g.][]{tg,sch,bl}, whereas the dependence of the variance parameter on the ensemble mean is similar to the expression of the variance in the gamma BMA model of \citet{srgf}, and it is in line with the relation of forecast uncertainty to the expected precipitation amount mentioned in the Introduction. Moreover, practical tests show that, at least for the UWME and ALADIN-HUNEPS ensemble considered in the case studies of Section \ref{sec:sec5}, models \eqref{eq:Link} and \eqref{eq:Linkex}, respectively, significantly outperform the corresponding CSG EMOS models with variance parameters 
\begin{equation*}
\sigma^2 = b_0+b_1 S^2 \qquad \text{and} \qquad \sigma^2 = b_0+b_1 \mathrm{MD},
\end{equation*}
where 
\begin{equation*}
 S^2:=\frac 1{m-1}\sum_{k=1}^m\big (f_k-\overline f\big)^2
\qquad \text{and} \qquad 
\mathrm{MD}:=\frac 1{m^2}\sum_{k,\ell=1}^m\big |f_k-f_{\ell}\big|
\end{equation*}
are the ensemble variance and the more robust ensemble mean difference \citep{sch}, respectively. Further, compared to the proposed models,
natural modifications
\begin{equation*}
\sigma^2 = b_0+b_1 S^2 +b_2\overline f\qquad \text{or} \qquad \sigma^2 = (b_0+b_1 \overline f)^2
\end{equation*}
in the CSG EMOS variance structure do not result in improved forecasts skills.

\subsection{Parameter estimation}
   \label{subs:subs2.2}

The main aim of probabilistic forecasting is to access the maximal sharpness of the predictive distribution subject to calibration \citep{gbr}. The latter means a statistical consistency between the predictive distributions and the validating observations, whereas the former refers to the concentration of the predictive distribution. This goal can be addressed with the help of scoring rules which measure the predictive performance by numerical values assigned to pairs of probabilistic forecasts and observations \citep{grjasa}. In atmospheric sciences the most popular scoring rules for evaluating predictive distributions are the logarithmic score, i.e. the negative logarithm of the predictive PDF evaluated at the verifying observation \citep{grjasa}, and the continuous ranked probability score \citep[CRPS;][]{grjasa,wilks}. For a
predictive CDF \ $F(y)$ \ and an observation \ $x$ \ the CRPS is defined as
\begin{equation}
  \label{eq:CRPS}
\crps\big(F,x\big):=\int_{-\infty}^{\infty}\big (F(y)-{\mathbbm 
  1}_{\{y \geq x\}}\big )^2{\mathrm d}y={\mathsf E}|X-x|-\frac 12
{\mathsf E}|X-X'|, 
\end{equation}
where \ ${\mathbbm 1}_H$ \ denotes the indicator of a set \ $H$, \ while \ $X$ \ and \ $X'$ \ are independent random variables with CDF \ $F$ \ and finite first moment. The CRPS can be expressed in the same units as the observation and one should also note that both scoring rules are proper \citep{grjasa} and negatively oriented, that is the smaller the better. 

For a CSG distribution defined by \eqref{eq:CSGCDF} the CRPS can be expressed in a closed form, \citet{schham} showed that
\begin{align*}
  \crps\big(G^{\,0}(k,\theta,\delta),x\big)=&\,(x+\delta)\Big(2G_{k,\delta} (x+\delta)-1\Big) -\frac{\theta k}{\pi}B\big(1/2, k+1/2\big)\Big(1-G_{2k,\delta}(2\delta)\Big)\\
&+\theta k\Big(1+2 G_{k,\delta}(\delta)G_{k+1,\delta}(\delta)-G_{k,\delta}^2(\delta) -2G_{k+1,\delta}(y+\delta)\Big)-\delta G_{k,\delta}^2(\delta).
\end{align*} 

Following the ideas of \citet{grwg} and \citet{sch}, the parameters of
models \eqref{eq:Link} (and \eqref{eq:Linkex} as well) are estimated
by minimizing the mean CRPS of predictive distributions and validating
observations corresponding to forecast cases of the training
period. We remark that optimization with respect to the mean
logarithmic score, that is, maximum likelihood (ML) estimation of
parameters, has also been investigated. Obviously, in terms of CRPS this model cannot outperform the one fit via CRPS minimization, however, in our test cases the ML method results in a reduction of the predictive skill of the CSG EMOS model in terms of almost all verification scores considered, so the corresponding values are not reported.

\section{Data}
  \label{sec:sec3}

\subsection{University of Washington mesoscale ensemble}
  \label{subs:subs3.1}

The eight-member UWME covers the Pacific Northwest region of North America and operates on a 12 km grid. The ensemble members are obtained from different runs of the fifth generation Pennsylvania State University--National Center for Atmospheric Research mesoscale model \citep[PSU-NCAR MM5;][]{grell} with initial and boundary conditions from various weather centres. We consider 48 h forecasts and corresponding validating observations of 24 h precipitation accumulation for 152 stations in the Automated Surface
Observing Network \citep{asos} in five US states. The forecasts are initialized at 0 UTC (5 PM local time when daylight saving time (DST) is in use and 4 PM otherwise) and we investigate data for calendar year 2008 with additional forecasts and observations from the last three months of 2007 used for parameter estimation. After removing days and locations with missing data 83 stations remain resulting in 20\,522 forecast cases for 2008.

\begin{figure}[t!]
\begin{center}
\leavevmode
\epsfig{file=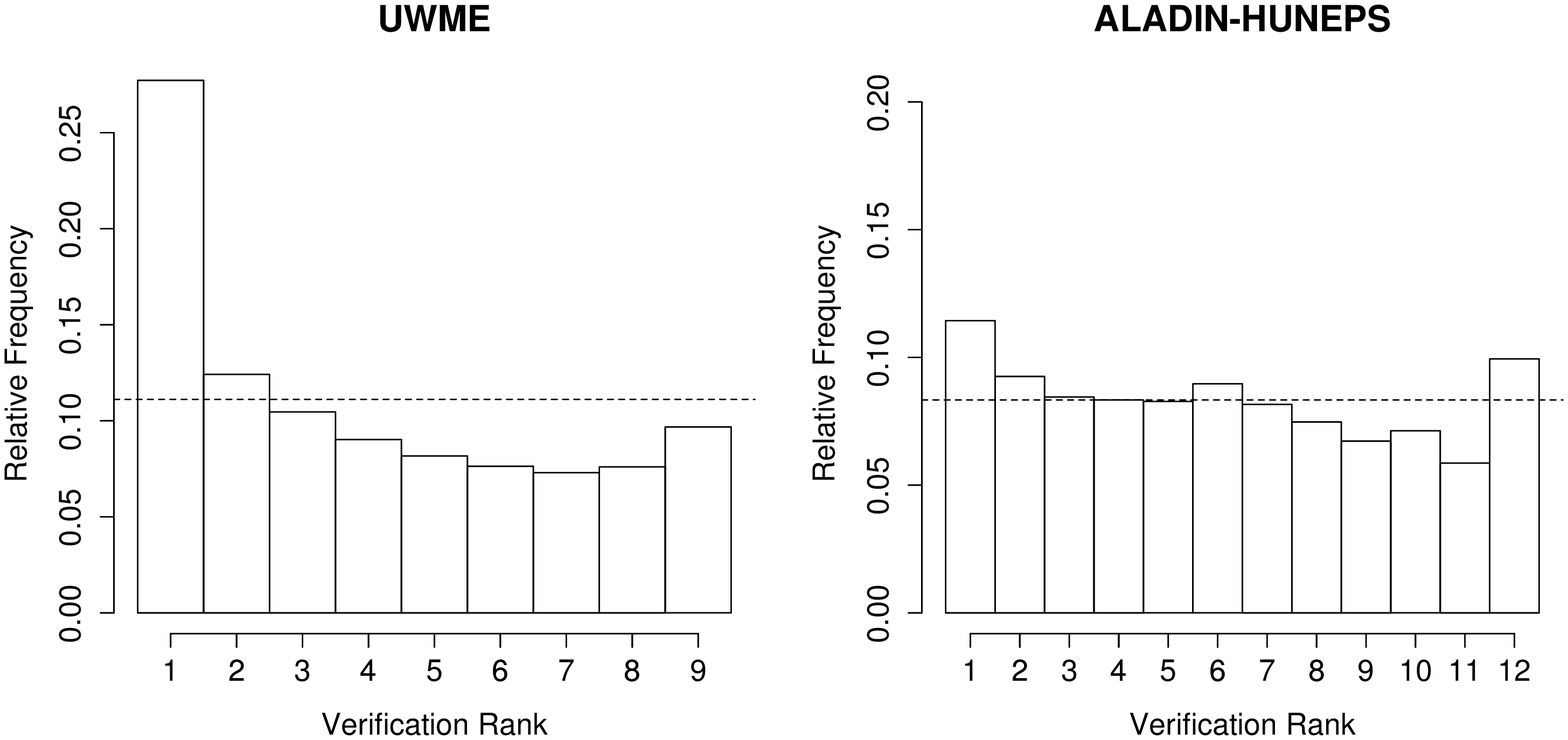,height=7cm,angle=0}

\centerline{\hbox to 9 cm{\small a) \hfill b)}}
\end{center}
\caption{Verification rank histograms. a) UWME for the calendar year 2008;    
  ALADIN-HUNEPS ensemble for the period  1 October 2010 -- 25 March 2011.} 
\label{fig:fig1}
\end{figure}

Figure \ref{fig:fig1}a shows the verification rank histogram of the raw ensemble, that is the histogram of ranks of validating observations with respect to the corresponding ensemble forecasts computed for all forecast cases \citep[see e.g.][Section 7.7.2]{wilks}, where zero observations are randomized among all zero forecasts. This histogram is far from the desired uniform distribution as in many cases the ensemble members overestimate the validating observation. The ensemble range contains the observed precipitation accumulation in $67.82\,\%$ of the cases, whereas the nominal coverage of the ensemble equals $7/9$, i.e $77.78\,\%$. Hence, the UWME is uncalibrated, and would require statistical post-processing to yield an improved forecast probability density function.

\subsection{ALADIN-HUNEPS ensemble}
   \label{subs:subs3.2}

The ensemble forecasts produced by the operational ALADIN-HUNEPS system of the HMS are obtained with dynamical downscaling of the global PEARP system of M\'et\'eo France by the ALADIN limited area model with an 8 km horizontal resolution. The EPS covers a large part of continental Europe and has 11 ensemble members, 10 exchangeable forecasts from perturbed initial conditions and one control member from the unperturbed analysis \citep{horanyi}.
The data base at hand contains ensembles of 42 h forecasts (initialized at 18 UTC, i.e. 8 pm local time when DST operates and 7 pm otherwise) for 24 h precipitation accumulation for 10 major cities in Hungary (Miskolc, Sopron, Szombathely, Gy\H or, Budapest, Debrecen, Ny\'\i regyh\'aza,
Nagykanizsa, P\'ecs, Szeged) together with the corresponding validating observations for the period between 1 October 2010 and 25 March 2011. The data set is fairly complete since there are only two dates when three ensemble members are missing for all sites. These dates are excluded from the analysis.

The verification rank histogram of the raw ensemble, displayed in Figure \ref{fig:fig1}b, shows far better calibration, than that of the UWME. The coverage of the ALADIN-HUNEPS ensemble equals $84.20\,\%$, which is very close to the nominal value of $83.33\,\%$ ($10/12$).

\section{Results}
  \label{sec:sec4}

As mentioned earlier, the predictive performance of the CSG EMOS model is tested on ensemble forecasts produced by the UWME and ALADIN-HUNEPS EPSs, and  
the results are compared with the fits of the GEV EMOS and gamma BMA models investigated by \citet{sch} and \citet{srgf}, respectively, and the verification scores of the raw ensemble. We remark that according to the suggestions of \citet{sch} for estimating the parameters of the GEV EMOS model for a given day, the estimates for the preceding day serve as initial conditions for the box constrained Broyden-Fletcher-Goldfarb-Shanno \citep{blnz} optimization algorithm. Compared with the case of fixed initial conditions this approach results in a slight increase of the forecast skills of the GEV EMOS model, whereas for the CSG EMOS method, at least in our case studies, fixed initial conditions are preferred. Further, we consider regional (or global) EMOS approach \citep[see e.g.][]{tg} which is based on ensemble forecasts and validating observations from all available stations during the rolling training period and consequently results in a single universal set of parameters across the entire ensemble domain.

\subsection{Diagnostics}
  \label{subs:subs4.1}

To get the first insight about the calibration of EMOS and BMA post-processed forecasts we consider probability integral transform (PIT) histograms. Generally, the PIT is the value of the predictive CDF evaluated at the verifying observation \citep{rgbp}, however, for our discrete-continuous models in the case of zero observed precipitation a random value is chosen uniformly from the interval between zero and the probability of no precipitation \citep{srgf}. Obviously, the closer the histogram to the uniform distribution, the better the calibration. In this way the PIT histogram is the continuous counterpart of the verification rank histogram  of the raw ensemble and provides a good measure about the possible improvements in calibration. 

The predictive performance of probabilistic forecasts is quantified with the help of the mean CRPS over all forecast cases, where for the raw ensemble the predictive CDF is replaced by the empirical one. Further, as suggested by \citet{gr}, Diebold-Mariano \citep[DM;][]{dm95} tests are applied for investigating the significance of the differences in scores corresponding to the various post-processing methods. The DM test takes into account the dependence in the forecasts errors and for this reason it is widely used in econometrics. 

\begin{figure}[t!]
\begin{center}
\leavevmode
\epsfig{file=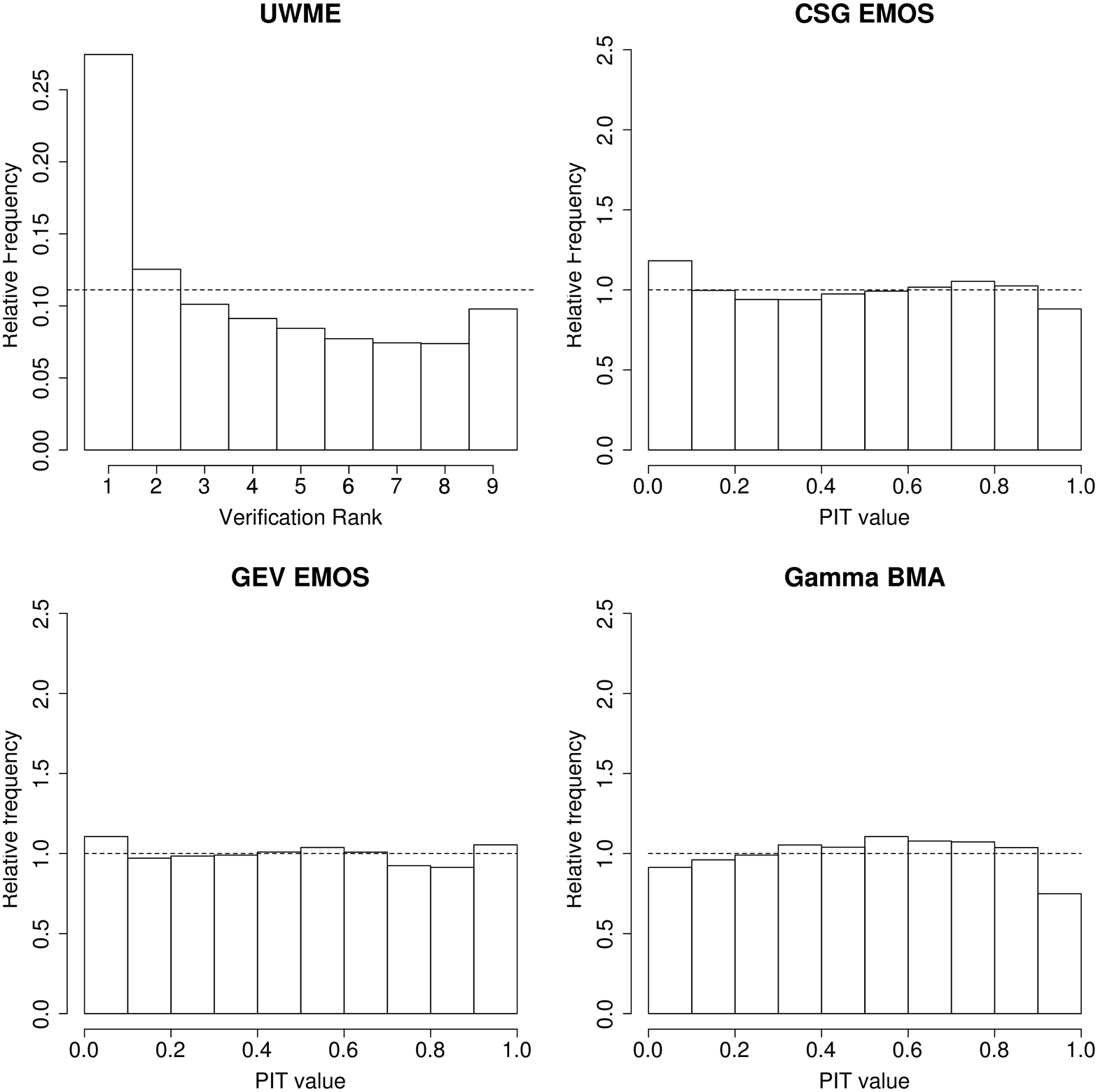,width=14cm,angle=0}
\end{center}
\caption{Verification rank histogram of the raw ensemble and PIT histograms of the EMOS and BMA post-processed forecasts for the UWME for the calendar year 2008.} 
\label{fig:fig2}
\end{figure}

Besides the CRPS we also consider Brier scores \citep[BS;][Section 8.4.2]{wilks} for the dichotomous event that the observed precipitation amount \ $x$ \ exceeds a given threshold \ $y$. \  For a predictive CDF \ $F(y)$ \ the probability of this event is \ $1-F(y)$, \ and the corresponding Brier score is given by
\begin{equation}
   \label{eq:eqBS}
 \bs \big(F,x;y\big):= \big (F(y)-{\mathbbm 1}_{\{y \geq x\}}\big )^2,
\end{equation}
see e.g. \citet{gr}. Obviously, the BS is negatively oriented and the CRPS \eqref{eq:CRPS} is the integral of the BSs over all possible thresholds. In our case studies we consider 0 mm precipitation, 5, 15, 25, 30 mm and 1, 5, 7, 9 mm threshold values for the UWME and ALADIN-HUNEPS ensemble, respectively, corresponding approximately to the 45th, 75th, 85th and 90th percentiles of the observed non-zero precipitation accumulations, and compare the mean BSs of the pairs of predictive CDFs and verifying observations over all forecast cases.

\begin{table}[t!]
\begin{center}{
\caption{$p$-values of Kolmogorov-Smirnov tests for uniformity
  of  PIT values for the UWME. Means of $10000$ random samples of sizes $2500$
  each.} \label{tab:tab1}  

  \vskip .5 truecm
\begin{tabular}{lccc} 
Model&CSG EMOS&GEV EMOS&Gamma BMA\\ \hline
Mean $p$-value&$0.154$&$0.310$&$0.044$
\end{tabular} }
\end{center}
\end{table}

The improvement in CRPS and BS with respect to a reference predictive distribution \ $F_{ref}$ \ can be measured with the help of the continuous ranked probability skill score (CRPSS) and the Brier skill score (BSS) defined as
\begin{equation*}
\crpss \big(F,x\big):=1-\frac{\crps \big(F,x\big)}{\crpss \big(F_{ref},x\big)}
\quad \text{and} \quad
\bss \big(F,x;y\big):=1-\frac{\bs \big(F,x;y\big)}{\bs \big(F_{ref},x;y\big)},
\end{equation*}
respectively \citep{wilks,ft12}. These scores are positively oriented and in our two case studies we use the raw ensemble as a reference.

To compare the calibration of probabilities of a dichotomous event of exceeding a given threshold calculated from the raw ensemble and the EMOS and BMA predictive distributions, we make use of reliability diagrams \citep[][Section 8.4.4]{wilks}. The reliability diagram plots the a graph of the observed frequency of the event against the binned forecast frequencies and in the ideal case this graph should lie on the main diagonal of the unit square.
In the case studies of Sections \ref{subs:subs4.1} and \ref{subs:subs4.2} we consider the same thresholds as for the BSs (UWME: 5, 15, 25, 30 mm; ALADIN-HUNEPS: 1, 5, 7, 9 mm;), whereas the unit interval is divided into 11 bins with break points \ $0.05,0.15,0.25,\ldots,0.95$. \ Following \citet{brsm} and \citet{sch}, the observed relative frequency of a bin is plotted against the mean of the corresponding probabilities, and we also add inset histograms displaying the frequencies of the different bins on $\log 10$ scales.

Further, one can investigate calibration and sharpness of a predictive distribution with the help of the coverage and average width of the \ $(1-\alpha )100\,\%, \ \alpha \in (0,1),$ \ central prediction interval. By coverage we mean the proportion of validating observations located between the lower and upper \ $\alpha /2$ \  quantiles of the predictive CDF and level
\ $\alpha$ \ should be chosen to match the nominal coverage of the raw ensemble, i.e. $77.78\,\%$ for the UWME and $ 83.33\,\%$ for the ALADIN-HUNEPS. As the coverage of a calibrated predictive distribution should be around \ $(1-\alpha )100\,\%$, \ the suggested choices of \ $\alpha$ \ allow direct comparisons with the raw ensembles, whereas the average width of the central prediction interval assesses the sharpness of the forecast.

Finally, point forecasts such as EMOS, BMA and ensemble medians are evaluated
with the help of mean absolute errors (MAEs) and DM tests for the forecast errors are applied to check whether the differences are significant.

\subsection{Verification results for the  UWME}
  \label{subs:subs4.2}

\begin{table}[t!]
\begin{center}
\caption{Mean CRPS of probabilistic forecasts, MAE of median forecasts and coverage and average width of $77.78\,\%$ central prediction intervals for the UWME.} \label{tab:tab2}

\vskip .5 truecm
\begin{tabular}{lcccc} 
Forecast&CRPS (mm)&MAE (mm)&Coverage (\%)&Av.width (mm) \\ \hline
CSG EMOS&2.252&3.019&80.46&8.350 \\
GEV EMOS&2.283&3.033&79.91&8.683 \\
Gamma BMA&2.357&3.220&83.44&9.515 \\
\hline
Ensemble&2.929&3.708&67.95&8.599 
\end{tabular}
\end{center}
\end{table}

The eight members of the UWME are generated using initial and boundary conditions from different sources, implying that the ensemble members are clearly distinguishable. Hence, the mean and the variance of the underlying gamma distribution of the CSG EMOS model are linked to the ensemble members according to \eqref{eq:Link} with \ $m=8$. \ Obviously, the reference GEV EMOS and gamma BMA models are also formulated under the assumption of non-exchangeable ensemble members. 

\begin{table}[t!]
\begin{center}
\caption{Values of the test statistics of the DM test for equal predictive performance based on the CRPS ({\em upper triangle}) and the prediction error of the median forecast ({\em lower triangle}) for the UWME. Negative/positive values indicate a superior predictive performance of the forecast given in the row/column label, bold numbers correspond to tests with $p$ values under $0.05$ level of significance. } \label{tab:tab3}

\vskip .5 truecm
\begin{tabular}{l|cccc} 
Forecast&CSG EMOS&GEV EMOS&Gamma BMA&Ensemble\\ \hline
CSG EMOS&--&\bf{-5.237}&\bf{-4.909}&\bf{-29.265} \\
GEV EMOS&1.631&--&\bf{-3.688}&\bf{-26.845} \\
Gamma BMA&{\bf 5.648}&{\bf 5.892}&--&\bf{-15.556} \\
Ensemble&{\bf 21.967}&{\bf 20.504}&{\bf 9.076}&-- 
\end{tabular}
\end{center}
\end{table}

A detailed study of CRPS and MAE values of the CSG EMOS and gamma BMA models corresponding to training period lengths of $20,\ 25, \ldots, 100$ days indicates that both scores have global minima at $70$ days.
Hence, in our analysis we calibrate the UWME forecasts for calendar year 2008 using this training period length.

\begin{table}[t!]
\begin{center}
\caption{CRPSS and BSS values with respect to the raw UWME.} \label{tab:tab4}

\vskip .5 truecm
\begin{tabular}{lcccccc} 
Forecast&CRPSS &\multicolumn{5}{c}{Brier Skill Score} \\ \cline{3-7}
& &0 mm&5 mm&15 mm&25 mm&30 mm \\ \hline
CSG EMOS&0.231&0.393&0.243&0.268&0.248&0.237 \\
GEV EMOS&0.221&0.403&0.219&0.252&0.239&0.235 \\
Gamma BMA&0.196&0.419&0.231&0.240&0.196&0.188 
\end{tabular}
\end{center}
\end{table}

Figure \ref{fig:fig2} showing the verification rank histogram of the raw ensemble and the PIT histograms of the CSG EMOS, GEV EMOS and gamma BMA models clearly illustrates the advantage of statistical post-processing. Unfortunately, the Kolmogorov-Smirnov (KS) test rejects the uniformity of the PIT values for all models, the highest $p$-value of \ $5.562\times 10^{-3}$ \ corresponds to the GEV EMOS approach. However, the small $p$-values are consequences of numerical problems caused by the large sample size \citep[see e.g.][]{bsv} and the mean $p$-values of $10000$ random samples of PITs of sizes $2500$ each, given in Table \ref{tab:tab1}, nicely follow the shapes of the histograms of Figure \ref{fig:fig2}.

\begin{figure}[t!]
\begin{center}
\leavevmode
\epsfig{file=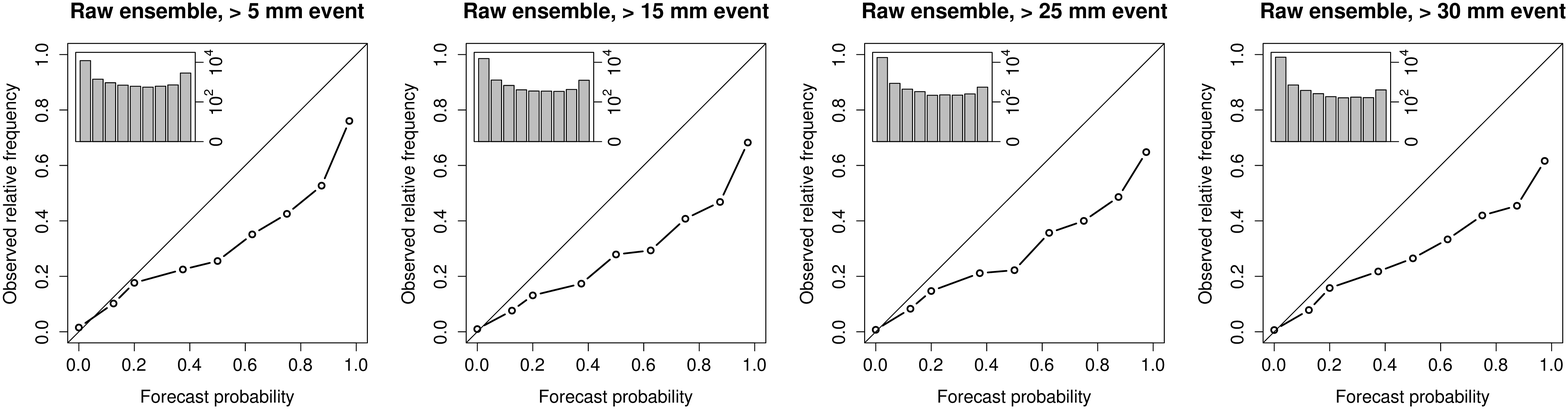,width=15.5cm,angle=0}

\vskip 2 mm

\epsfig{file=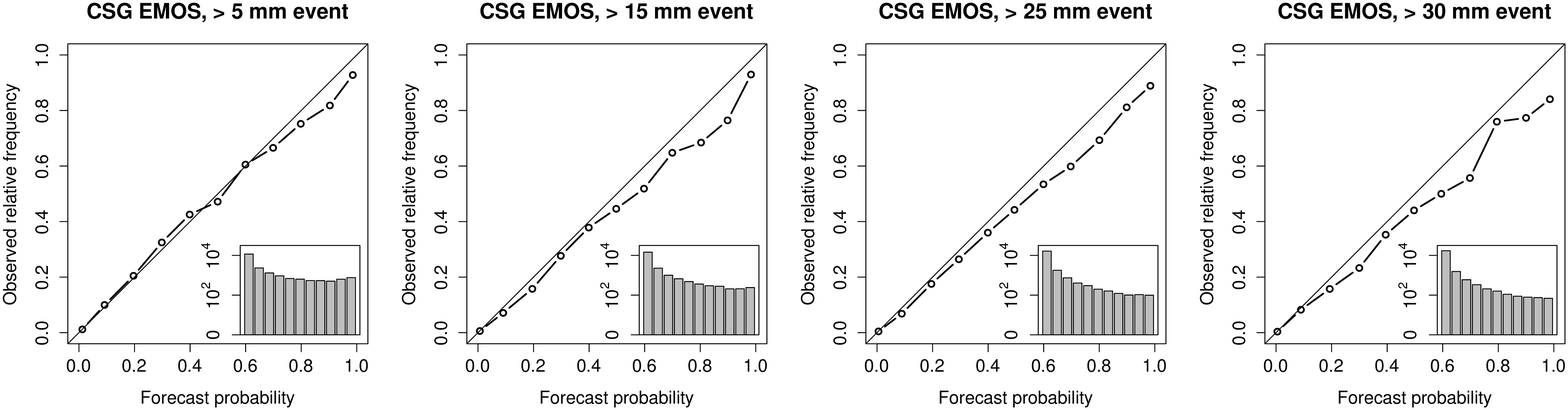,width=15.5cm,angle=0}

\vskip 2 mm

\epsfig{file=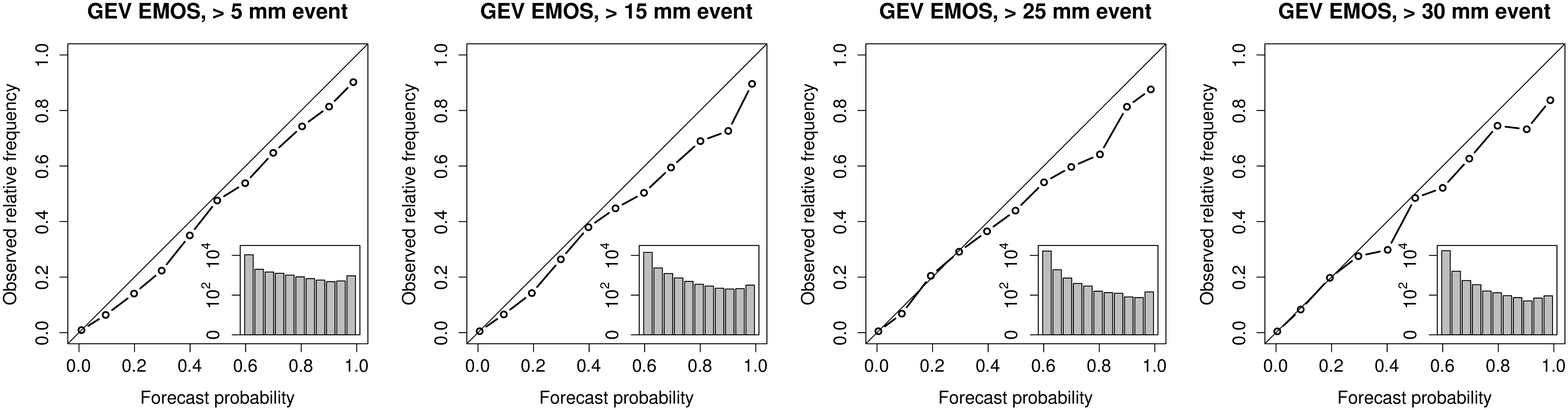,width=15.5cm,angle=0}

\vskip 2 mm

\epsfig{file=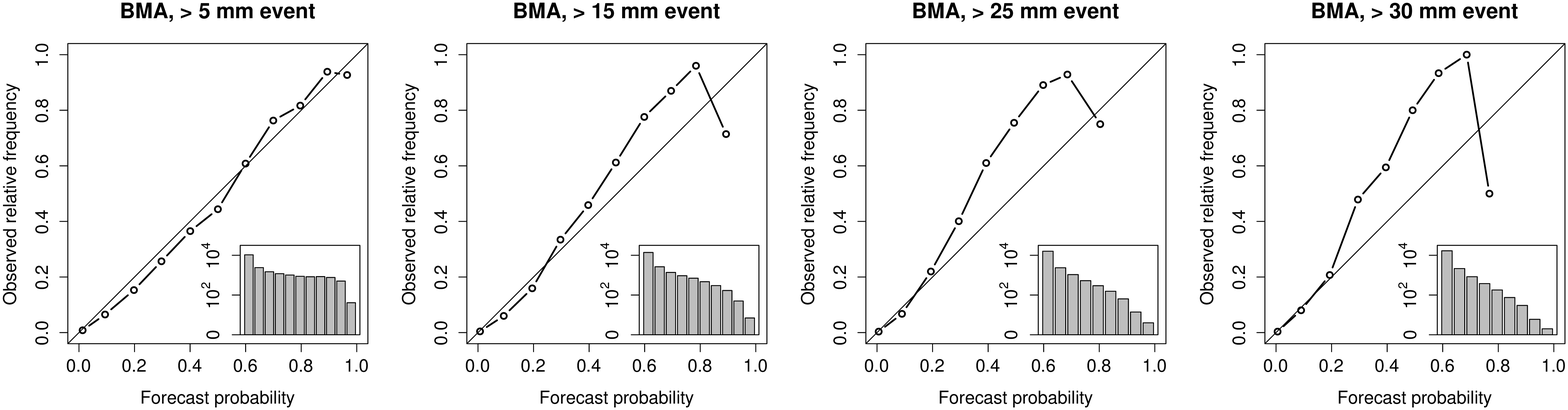,width=15.5cm,angle=0}
\end{center}
\caption{Reliability diagrams of the raw ensemble and EMOS and BMA post-processed forecasts for the UWME for the calendar year 2008.
The inset histograms display the log-frequency of cases within the respective bins.} 
\label{fig:fig3}
\end{figure}

In Table \ref{tab:tab2} the mean CRPS of probabilistic forecasts, the MAE of median forecasts and the coverage and average width of $77.78\,\%$ central prediction intervals for the two EMOS approaches, the gamma BMA model and the raw ensemble are reported, whereas Table \ref{tab:tab3} shows the results of DM tests for equal predictive performance based on the CRPS values and the prediction errors of the median. 
By examining these results, one can clearly observe the obvious advantage of post-processing with respect to the raw ensemble, which is quantified in the significant decrease of CRPS and MAE values and in a substantial improvement in coverage. Further, the CSG EMOS model results in the lowest CRPS value, whereas in terms MAE there is no difference between the two EMOS methods, which significantly outperform the gamma BMA approach both in calibration of probabilistic and accuracy of point forecasts. The CSG EMOS model results in the sharpest central prediction interval combined with a rather fair coverage, whereas the central prediction intervals corresponding to the other two calibration methods are slightly wider than that of the raw ensemble. 

\begin{figure}[t!]
\begin{center}
\leavevmode
\epsfig{file=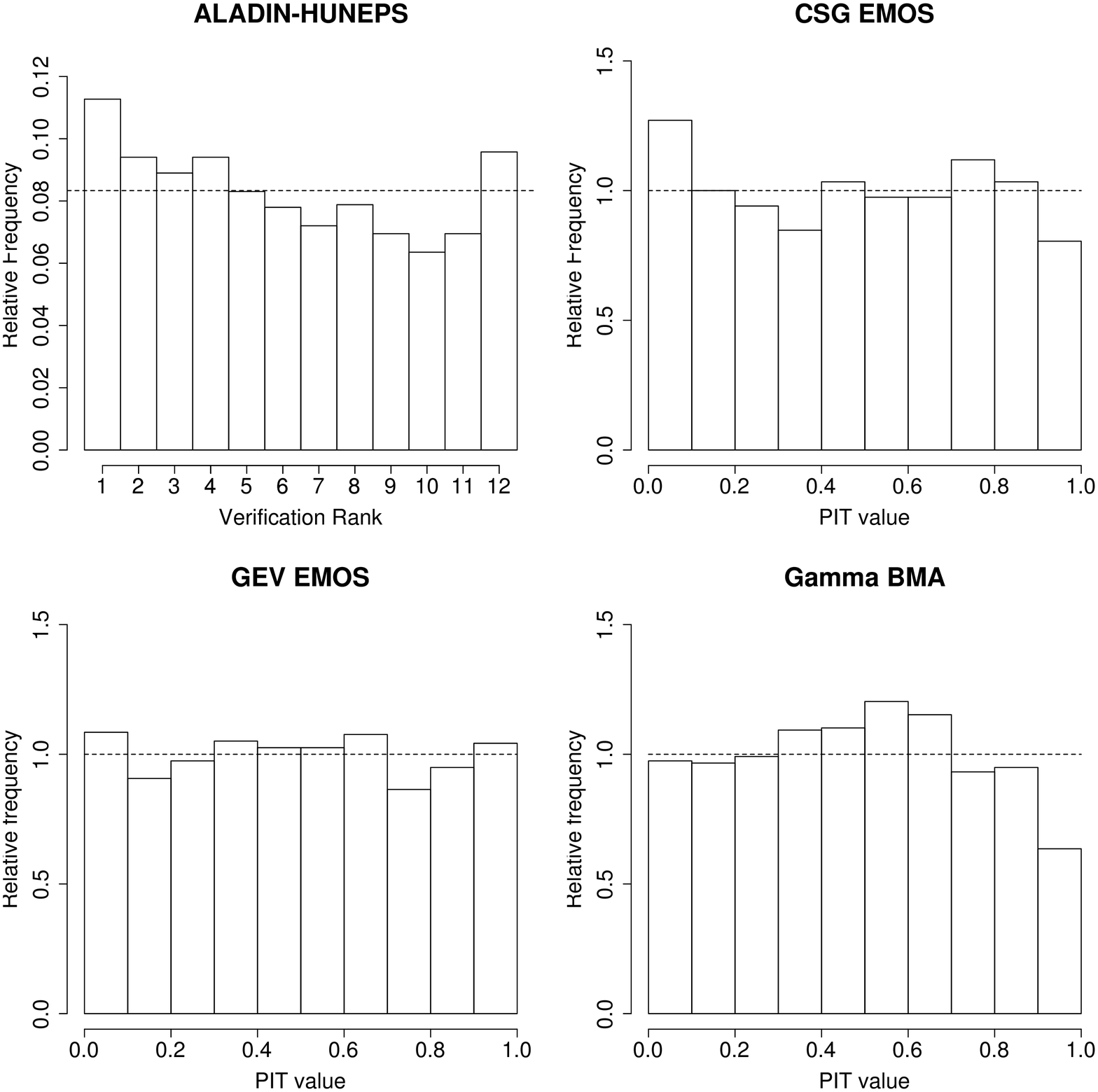,width=14cm,angle=0}
\end{center}
\caption{Verification rank histogram of the raw ensemble and PIT histograms of the EMOS and BMA post-processed forecasts for the  ALADIN-HUNEPS ensemble for the period 27 November 2010 -- 25 March 2011.} 
\label{fig:fig4}
\end{figure}

The improvement in calibration caused by statistical post-processing can also be observed in skill scores reported in Table \ref{tab:tab4} and reliability diagrams displayed in Figure \ref{fig:fig3}. Gamma BMA method performs well in predicting the probability of positive precipitation and exceeding the 5 mm threshold, whereas for higher threshold values it is behind the two EMOS approaches, where the CSG EMOS model presents slightly better forecast skills. Hence, one can conclude, that in case of the UWME the EMOS approaches outperform both the raw ensemble and the gamma BMA model and the proposed CSG EMOS model slightly outperforms the GEV EMOS method.

\subsection{Verification results for the ALADIN-HUNEPS ensemble}
   \label{subs:subs4.23}

\begin{table}[t!]
\begin{center}
\caption{$p$-values of Kolmogorov-Smirnov tests for uniformity
  of  PIT values for the ALADIN-HUNEPS ensemble.} \label{tab:tab5}  

\vskip .5 truecm 
\begin{tabular}{lccc} 
Model&CSG EMOS&GEV EMOS&Gamma BMA\\ \hline
$p$-value&$0.119$&$0.921$&$0.003$
\end{tabular} 
\end{center}
\end{table}

As a contrast to the UWME, the way the ALADIN-HUNEPS ensemble is generated (see Section \ref{subs:subs3.2}) induces a natural grouping of the ensemble members. The first group contains the control, whereas the second group consists of the 10 exchangeable ensemble members. This splitting results in the GEV EMOS model 
\eqref{eq:Linkex} with \ $m=2$, \ $M_1=1$ \ and \ $M_2=10$ and the same grouping is considered for the benchmarking GEV EMOS and gamma BMA models \citep{frg}.

\begin{table}[t!]
\begin{center}
\caption{Mean CRPS of probabilistic forecasts, MAE of median forecasts and coverage and average width of $83.33\,\%$ central prediction intervals for the ALADIN-HUNEPS ensemble.} \label{tab:tab6}

\vskip .5 truecm
\begin{tabular}{lcccc} 
Forecast&CRPS (mm)&MAE (mm)&Coverage (\%)&Av.width (mm) \\ \hline
CSG EMOS&0.465&0.636&89.15&2.185 \\
GEV EMOS&0.477&0.641&86.53&2.192 \\
Gamma BMA&0.532&0.708&93.73&2.854 \\
\hline
Ensemble&0.485&0.640&84.24&2.436 
\end{tabular}
\end{center}
\end{table}

Again, in order to determine the appropriate length of the rolling training period the mean CRPS and MAE values of the various models for training periods of lengths $20,\ 25, \ldots, 100$ calendar days are investigated. In order to ensure the comparability of the results corresponding to different training period lengths, verification scores from 10 January to 25 March 2011 are considered. The corresponding curves of the CRPS and MAE scores plotted against the training period lengths (not shown) have global minima at $85$ days, however they have elbows at $55$ days, that is, up to this training period length the decrease is rather steep then the values stabilize. Hence, as in general shorter training periods are preferred, for calibrating the ALADIN-HUNEPS ensemble a training period of length $55$ days is used. This means that ensemble members, validating observations, and predictive PDFs are available for the period from 27 November 2010 to 25 March 2011 having 119 calendar days (just after the first $55$ day
training period) and 1180 forecast cases, since on 15 February 2011 three ensemble members are missing and this date is excluded from the analysis.   This time interval starts more than 6  weeks earlier than the one
used for determination of the optimal training period length.

\begin{table}[t!]
\begin{center}
\caption{Values of the test statistics of the DM test for equal predictive performance based on the CRPS ({\em upper triangle}) and the prediction error of the median forecast ({\em lower triangle}) for the ALADIN-HUNEPS ensemble. Negative/positive values indicate a superior predictive performance of the forecast given in the row/column label, bold numbers correspond to tests with $p$ values under $0.05$ level of significance. } \label{tab:tab7}

\vskip .5 truecm 
\begin{tabular}{l|cccc} 
Forecast&CSG EMOS&GEV EMOS&Gamma BMA&Ensemble\\ \hline
CSG EMOS&--&\bf{-2.758}&\bf{-3.978}&\bf{-2.928} \\
GEV EMOS&0.799&--&\bf{-3.586}&-1.177 \\
Gamma BMA&{\bf 2.682}&{\bf 2.697}&--&{\bf 2.498} \\
Ensemble&0.246&-0.078&{\bf -2.109}&--
\end{tabular}
\end{center}
\end{table}

Compared with the verification rank histogram of the raw ensemble the PIT histograms of the post-processed forecasts displayed in Figure \ref{fig:fig4} show a substantial improvement in calibration. For the two EMOS models the KS test accepts the uniformity of the PIT values (see Table \ref{tab:tab5} and note the extremely high $p$-value for the GEV EMOS), whereas the histogram of the Gamma BMA model is hump shaped indicating some overdispersion.

\begin{table}[t!]
\begin{center}
\caption{CRPSS and BSS values with respect to the raw ALADIN-HUNEPS ensemble.} \label{tab:tab8}

\vskip .5 truecm 
\begin{tabular}{lrrrrrr} 
Forecast&CRPSS &\multicolumn{5}{c}{Brier Skill Score} \\ \cline{3-7}
& &0 mm&1 mm&5 mm&7 mm&9 mm \\ \hline
CSG EMOS&0.042&0.094&0.057&-0.011&-0.025&0.019\\
GEV EMOS&0.017&0.166&0.008&-0.022&-0.030&0.027\\
Gamma BMA&-0.098&0.151&-0.070&-0.265&-0.136&-0.023
\end{tabular}
\end{center}
\end{table}

Concerning the two EMOS approaches, the verification scores of Table \ref{tab:tab6} together with the results of the corresponding DM tests for equal predictive performance (see Table \ref{tab:tab7}) display similar behavior as in the case of the UWME. There is no significant difference between the MAE values of the CSG and GEV EMOS methods and the former results in the lowest CRPS and the sharpest $83.33\,\%$ central prediction interval. Further, the EMOS models significantly outperform both the raw ensemble and the gamma BMA approach, despite the raw ensemble is rather well calibrated and has far better predictive skill than the BMA calibrated forecast. Note that the large mean CRPS and coverage of the BMA predictive distribution is totally in line with the shape of the corresponding PIT histogram of Figure \ref{fig:fig4}.

\begin{figure}[t!]
\begin{center}
\leavevmode
\epsfig{file=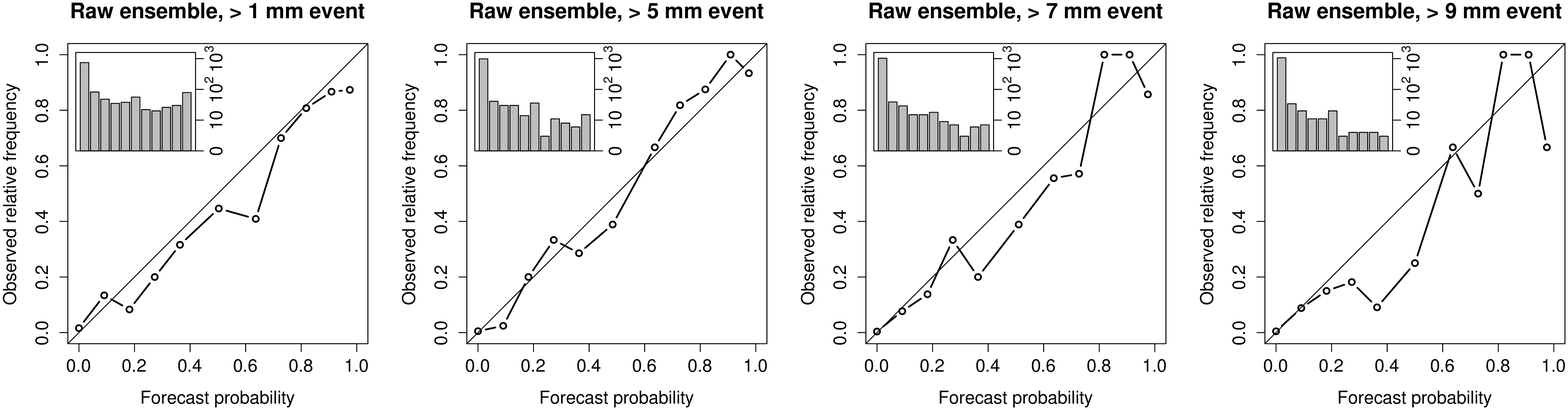,width=15.5cm,angle=0}

\vskip 2 mm

\epsfig{file=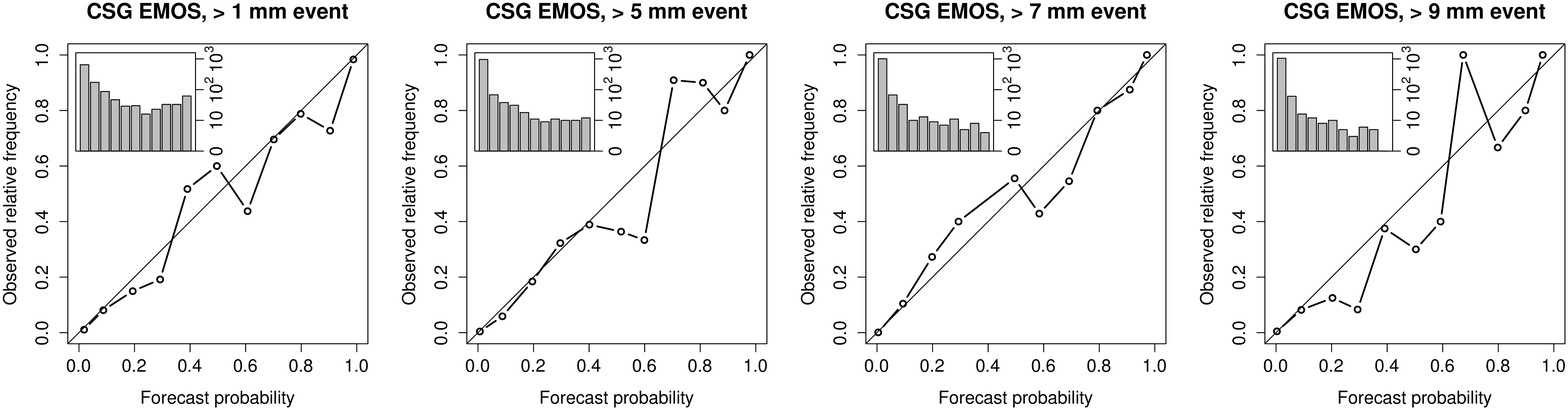,width=15.5cm,angle=0}

\vskip 2 mm

\epsfig{file=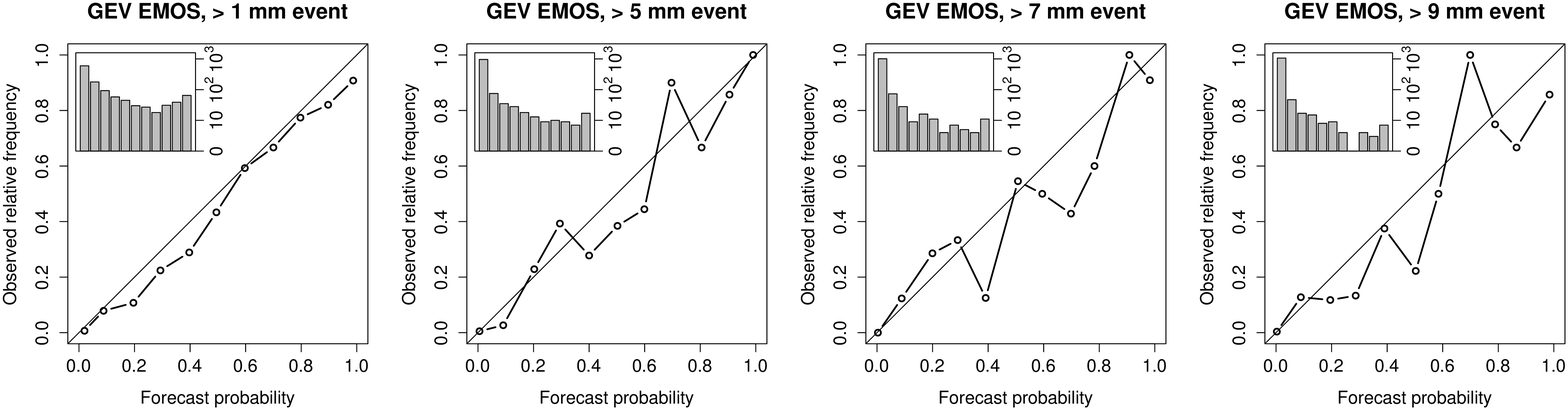,width=15.5cm,angle=0}

\vskip 2 mm

\epsfig{file=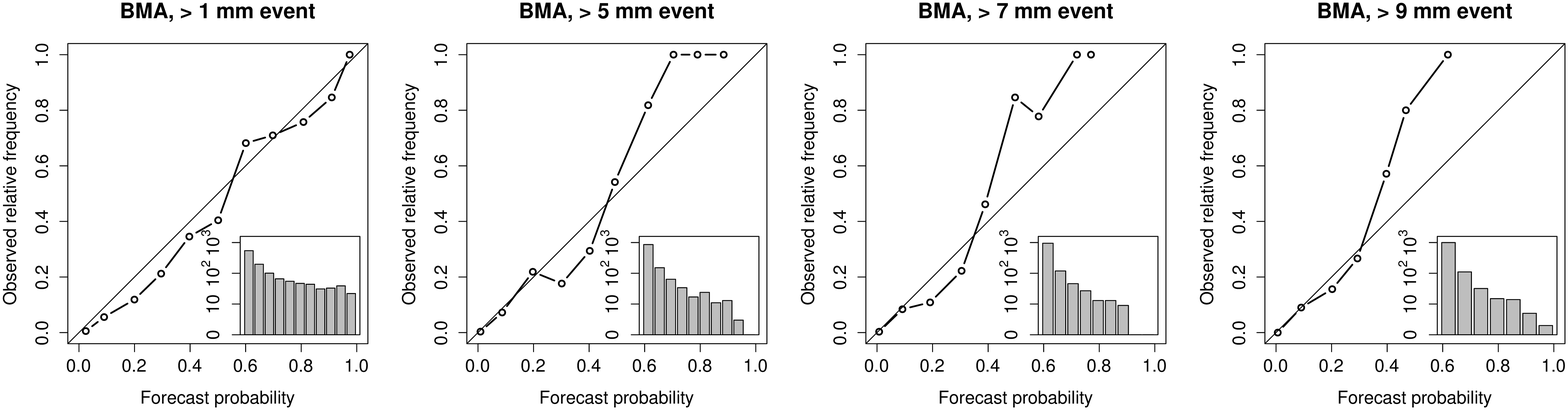,width=15.5cm,angle=0}
\end{center}
\caption{Reliability diagrams of the raw ensemble and EMOS and BMA post-processed forecasts for the  ALADIN-HUNEPS ensemble for the period 27 November 2010 -- 25 March 2011. The inset histograms display the log-frequency of cases within the respective bins.} 
\label{fig:fig5}
\end{figure}

The good predictive performance of the ALADIN-HUNEPS ensemble can also be observed on the large amount of negative skill scores reported in Table \ref{tab:tab8} and on the reliability diagrams of Figure \ref{fig:fig5}. Similar to the case of the UWME, for 0 mm threshold the gamma BMA model has good predictive performance, whereas for higher threshold values it underperforms the CSG and GEV EMOS models and the raw ensemble.  However, in connection with the reliability diagrams one should also note that the hectic behavior of the graphs (compared to the rather smooth diagrams of Figure \ref{fig:fig3}) is a consequence of the shortage of data, as the verification period contains only 394 observations of positive precipitation, which is around one third of the forecast cases.

Taking into account both the uniformity of the PIT values and the verification scores in Tables \ref{tab:tab6} and \ref{tab:tab8} it can be said that the proposed CSG EMOS model has the best overall performance in calibration of the raw ALADIN-HUNEPS ensemble forecasts of precipitation accumulation.

\section{Conclusions}
  \label{sec:sec5}

A new EMOS model for calibrating  ensemble forecasts of precipitation accumulation is proposed where the predictive distribution follows a censored and shifted gamma distribution, with  mean and variance of the underlying gamma law being affine functions of the raw ensemble and the ensemble mean, respectively. The CSG EMOS method is tested on ensemble forecasts of 24 h precipitation accumulation of the eight-member University of Washington mesoscale ensemble and on the 11 member ALADIN-HUNEPS ensemble of the Hungarian Meteorological Service. These ensemble prediction systems differ both in the climate of the covered area and in the generation of the ensemble members. 
By investigating the uniformity of the PIT values of predictive distributions, the mean CRPS of probabilistic forecasts, the Brier scores and reliability diagrams for various thresholds, the MAE of median forecasts and the average width and coverage of central prediction intervals corresponding to the nominal coverage, the predictive skill of the new approach is compared with that of the GEV EMOS method \citep{sch}, the gamma BMA model \citep{srgf} and the raw ensemble.
From the results of the presented case studies one can conclude that in terms of calibration of probabilistic and accuracy of point forecasts the proposed CSG EMOS model significantly outperforms both the raw ensemble and the BMA model and shows slightly better forecast skill than the GEV EMOS approach.

\bigskip
\noindent
{\bf Acknowledgments.} \ 
S\'andor Baran is supported by
the J\'anos Bolyai Research Scholarship of the Hungarian Academy of Sciences.
D\'ora Nemoda partially carried out her research in the framework of the Center of Excellence of Mechatronics and Logistics  at the University of Miskolc. The authors are indebted to Michael Scheuerer for his useful suggestions and remarks and for providing the {\tt R} code for the GEV EMOS model. The authors further thank the University of Washington MURI group for providing the UWME data and Mih\'aly Sz\H ucs from the HMS for the ALADIN-HUNEPS data.

\end{document}